\documentclass{elsart}
\usepackage{amsfonts}
\usepackage{graphicx}
\usepackage{amssymb}

\newcommand{\wb}{\omega_{\mathrm{b}}}

\newcommand{\middlefig}{.5\textwidth}
\newcommand{\singlefig}{.5\textwidth}

\begin{document}

\begin{frontmatter}

\title{Interaction of moving discrete breathers with vacancies}

\author{J Cuevas,\thanksref{cor}}
\author{JFR Archilla,}
\author{B S\'{a}nchez-Rey,}
\author{FR Romero}
\address{Grupo de F\'{\i}sica No Lineal (Nonlinear Physics Group).
Universidad de Sevilla. Avda. Reina Mercedes, s/n. 41012-Sevilla
(Spain)}
\thanks[cor]{Corresponding author. E-mail: jcuevas@us.es}


\date{14 October 2005}

\journal{Physica D}

\begin{keyword}
Discrete breathers \sep Mobile breathers \sep Intrinsic localized
modes \sep Vacancies \sep Breather-kink interaction.

\PACS
63.20.Pw,  
\sep
63.20.Ry,  
\sep
63.50.+x, 
\sep
66.90.+r 
\end{keyword}

\begin{abstract}
In this paper a Frenkel--Kontorova model with a nonlinear interaction potential
is used to describe a vacancy defect in a crystal. According to recent
numerical results [Cuevas \emph{et al}. Phys. Lett. A 315, 364 (2003)] the
vacancy can migrate when it interacts with a moving breather. We study more
thoroughly the phenomenology caused by the interaction of moving breathers with
a single vacancy and also with  double vacancies. We show that vacancy mobility
is strongly correlated with the existence and stability properties of
stationary breathers centered at the particles adjacent to the vacancy, which
we will now call vacancy breathers.
\end{abstract}

\end{frontmatter}

\section{Introduction}

Discrete breathers (DBs) are classical, spatially localized,
time-periodic, numerically exact solutions which can be sustained by
many non-linear lattices~\cite{FW98}. Their existence, which is
proven by rigorous theorems~\cite{MA94} and a large amount of
numerical results, is not restricted to special or integrable
models. On the contrary, they can be found, in principle, in any
discrete, nonlinear system and in any dimension
\cite{PHYSD99,CHAOS03}. They have been observed in experiments
involving different systems, as Josephson-junctions arrays
\cite{BAUFZ00,TMO00}, waveguide arrays \cite{Eis,FCSEC03}, molecular
crystals \cite{Swa} and antiferromagnetic systems \cite{SES99}. They
are also thought to play an important role in DNA denaturation
\cite{P04}.

Usually they are pinned to the lattice but under certain circumstances may
become highly mobile \cite{CAT96}. In this case, an interesting problem arises:
the interaction between moving discrete breathers (MBs) and local
inhomogeneities. This problem has been addressed within different frameworks:
scattering of breathers or solitons by impurities~\cite{CPAR02b,FPM94,BK98}, by
lattice junctions~\cite{BSS02} and by bending points~\cite{TSI02} of a given
chain.

Vacancies in a crystal are another type of local inhomogeneities. By using
numerical methods, some of the authors found that a localized energy packet in
the form of a moving discrete breather can put a vacancy into
movement~\cite{CKAER03}. This vacancy migration induced by localized
excitations had already been suggested as an explanation for some experimental
results~\cite{SAR00}.

The aim of this paper is to find an explanation to the somehow
qualitative results established in \cite{CKAER03}. The main result
is that vacancies mobility is highly dependent on the existence and
stability of the breathers adjacent to the vacancy, hereafter called
vacancy breathers (VBs).

\section{The model}

The simplest way to model a vacancy in a crystal consists in
considering a one-dimensional chain of interacting particles
submitted to a periodic substrate potential. It is known in other
context as the Frenkel-Kontorova (FK) model~\cite{FK38}. A vacancy
can be represented by an empty well of the substrate potential as
Fig.~1 shows.

\begin{figure}
\begin{center}
    \includegraphics[width=\singlefig]{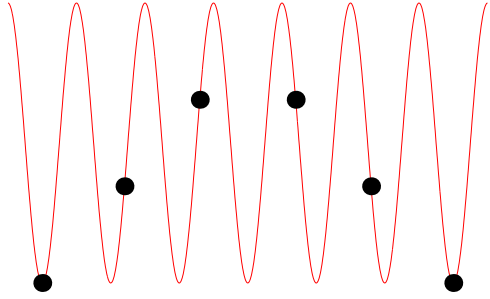} 
\caption{Scheme of the equilibrium state of the Frenkel--Kontorova
model with sine-Gordon substrate potential and Morse nearest
neighbor interaction. The empty well represents a vacancy.}
\label{fig:FK}
\end{center}
\end{figure}

 The  Hamiltonian of this system is~\cite{BK98}
\begin{equation}
    H=\sum_{n=1}^N\frac{1}{2} m\dot
    x_n^2+V(x_n)+ W(x_n-x_{n-1}) \quad ,
\end{equation}
being $x_n$  the absolute coordinate of the $n$-th particle. We
have chosen the sine-Gordon potential \begin{equation}
V(x_n)=\frac{a^2}{4\pi^2}[1-\cos(2\pi x_n/a)]\end{equation} as the
simplest periodic, substrate potential, with linear frequency
normalized to the unity $\omega_0=\sqrt{V''(0)}=1$.  We have
chosen the Morse potential for the interaction between particles
because its force weakens as the distance between particles grows
and its hard part prevents the particles from crossing. It is
given by
\begin{equation}
    W(x_n-x_{n-1})=\frac{C}{2b^2} [e^{-b(x_n-x_{n-1}-a)}-1]^2.
\end{equation}%
The parameter $C=W''(a)$ is the curvature of the Morse potential.
We have taken $C=0.5$ in order that MBs exist in this system for a
breather frequency $\omega_b=0.9$. The strength of the interaction
potential can be modulated, without changing its curvature, by
varying $b$, being $b^{-1}$  a measure of the well width and
$C/2b^2$ the well depth. We have also normalized the lattice
period $a$ and the masses to the unity. The dynamical equations
are given by
\begin{equation}
\ddot x_n+
   V'(x_n)+[W'(x_n-x_{n-1})-W'(x_{n+1}-x_n)]=0 .
   \label{eq:dyn}
\end{equation}

In this system, it is possible to generate DBs numerically using the
standard methods from the anticontinuous limit~\cite{MA96}. We can
also induce translational motion of DBs by using a simplified form
of the marginal mode method~\cite{CAT96}. It consists of adding  a
perturbation $\vec{v}=\lambda
(...,0,-1/\sqrt{2},0,1/\sqrt{2},0,...)$ to the velocities of the
stationary breather, with the nonzero values at the neighboring
sites of the initial breather center. The resulting DB kinetics is
very smooth and resembles that of a classical free particle with
constant velocity. Therefore, we can consider the total energy of a
MB as the sum of a the internal energy, equal to the one of the
stationary breather, plus the translational energy,  equal to the
energy of the perturbation added $K=\lambda^2/2$.

In order to facilitate systematic studies, we have introduced
dissipation at the boundaries.

\section{Interaction of moving breathers with a single vacancy}

\begin{figure}
\begin{center}
\begin{tabular}{cc}
    \includegraphics[width=\middlefig]{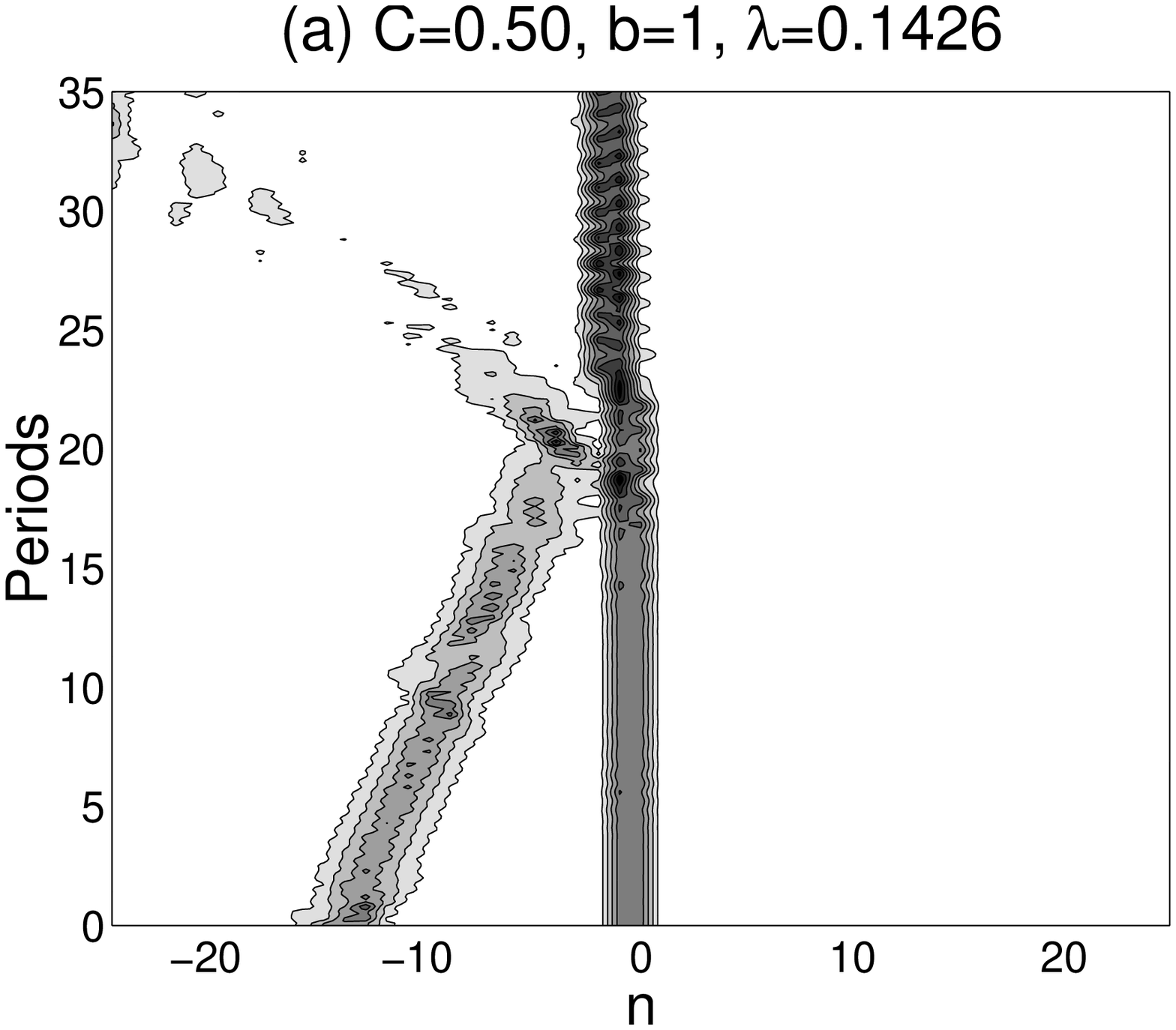} &
    \includegraphics[width=\middlefig]{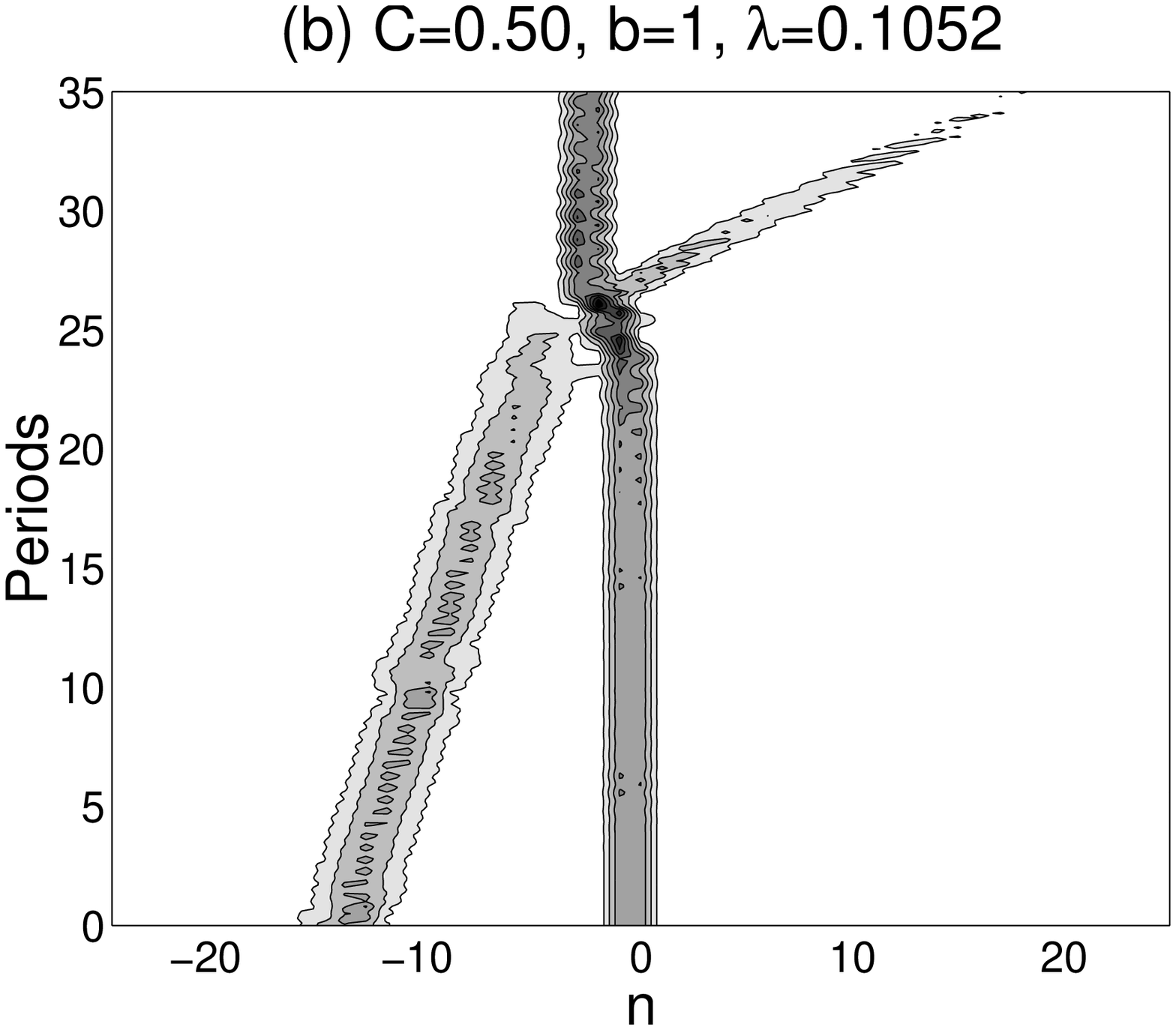}\\
    \includegraphics[width=\middlefig]{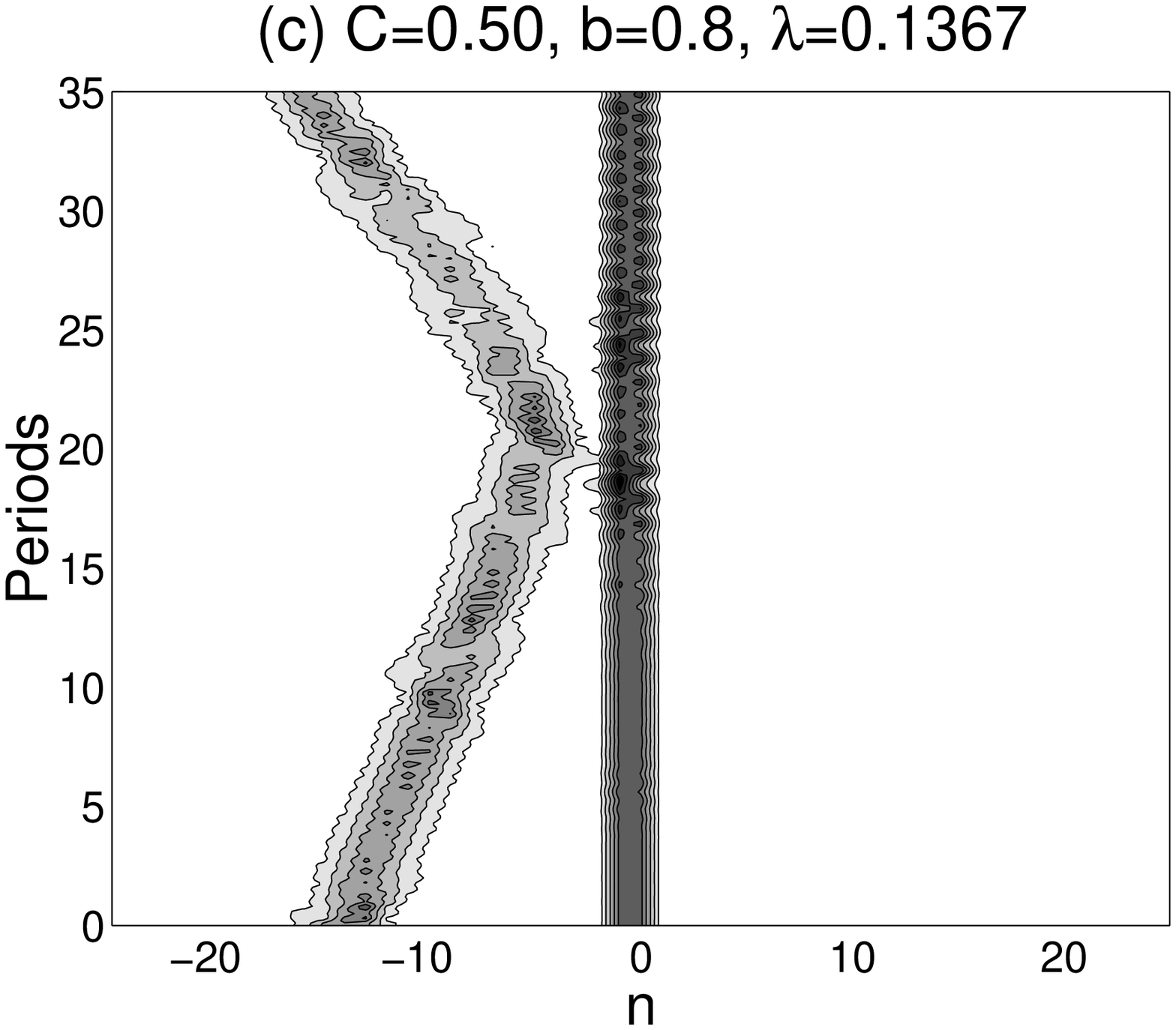} &
    \includegraphics[width=\middlefig]{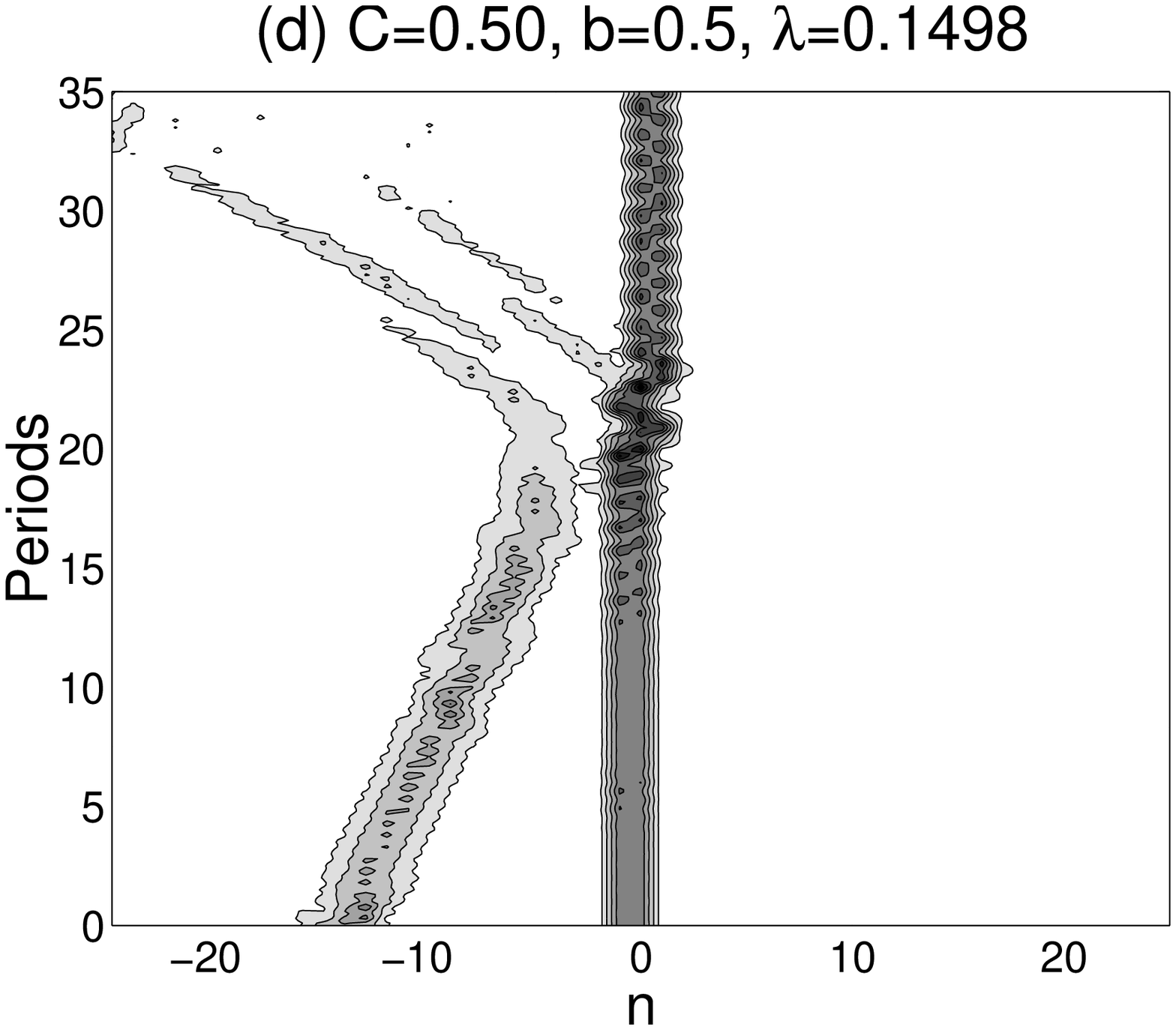}
\end{tabular}
\caption{Energy density plot for the interaction moving breather--vacancy. The
particle to the right of the vacancy is located at $n=0$. In (a), (c) and (d)
the moving breather is reflected and the vacancy moves backwards, remains at
rest and moves forwards respectively.  In (b) the breather is transmitted and
the vacancy moves backwards.} \label{fig:edp1v}
\end{center}
\end{figure}

In order to investigate vacancy mobility, we have generated DBs far from a
vacancy and then launched this breather against it. Our numerical calculations
show that the outcome of the scattering is extremely sensitive to the initial
conditions~\cite{CKAER03}. The incident breather can be reflected, trapped or
transmitted (see Fig.~\ref{fig:edp1v}), always losing energy as it occurs in
the interaction between a MB and an impurity~\cite{CPAR02b}, whereas the
vacancy can either move forward or backward or remain at rest. This scenario is
very different to the one arising in the continuum limit where the vacancy
(anti-kink) always moves backwards and the breather is always transmitted after
the collision.

We have studied the dependence of the vacancy mobility on the
parameter $b$, which controls the strength of the interaction
between neighboring particles of the chain. For each value of $b$
we have analyzed a set of 601 numerical experiments of scattering
corresponding to different initial conditions. As initial
conditions we have considered a set of MBs with increasing kinetic
energy, which were obtained choosing the values of the parameter
$\lambda$ uniformly distributed in the interval $[0.10,0.16]$. The
lower bound of this distribution is chosen so that the threshold
value for vacancies movement (see below) is surpassed. The upper
bound can be increased without any dramatic change in the results.
However, an extremely high value of $\lambda$ can destroy the
moving breather.

\begin{figure}
\begin{center}
\includegraphics[width=\singlefig]{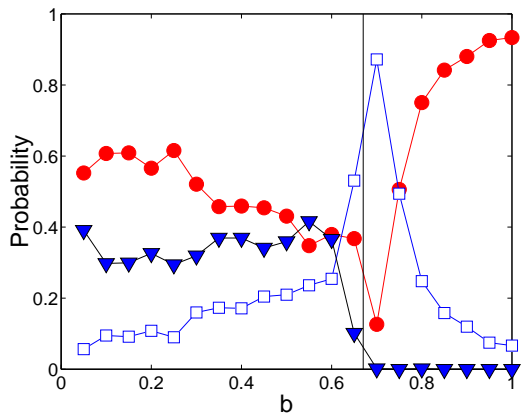} 
\caption{Probability for the vacancy to remain at its site
(squares), move backward (circles) or forward (triangles). The
interaction weakens when $b$ increases. Note that no forward
movement of the vacancy occurs to the right of the vertical line.
\label{fig:prob}}
\end{center}
\end{figure}

Fig.~\ref{fig:prob} shows the probability that the vacancy moves
forward, backward or remains at rest (averaged over the set of
initial conditions) with respect to the parameter $b$. Note that,
as it should be expected, the backward movement is the most
probable behavior because in this case the incident breather
pushes forward the particle to the left of the vacancy. In fact,
the forward movement of the vacancy requires a strong enough
interaction, i.e. $b$ has to be smaller that a critical value
$b_f\approx 0.7$.

\begin{figure}
\begin{center}
    \includegraphics[width=\singlefig]{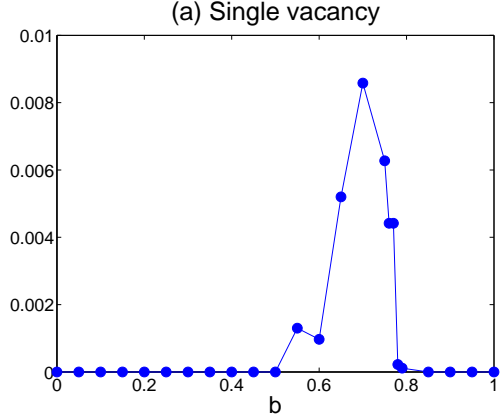} 
\caption{Minimal translational energy ($K_{min}$) to move a
vacancy.} \label{fig:Kmin} \end{center}
\end{figure}

Another interesting finding is shown in Fig.~\ref{fig:Kmin}: the
translational energy of the MB has to be higher than a minimal
value $K_{min}$ in a certain interval $b\in (0.5,0.8)$ in order to
move the vacancy. It is worth remarking that $K_{min}=0$ means the
minimum kinetic energy needed to move a breather, which is
actually different to zero. However, we have included this
notation in order to have a clear picture of the vacancy movement.

\section{Vacancy modes and vacancy breathers}

We can try to understand these results by studying: a)~the linear
modes, which can be obtained by linearizing Eqs.~\ref{eq:dyn}
around the equilibrium state shown in Fig.~1; and b)~the vacancy
breathers, i.e., the  nonlinear localized modes centered at the
particles adjacent to the vacancy, with the same frequency as the
incident MB.

\begin{figure}
\begin{center}
\begin{tabular}{cc}
    \includegraphics[width=\middlefig]{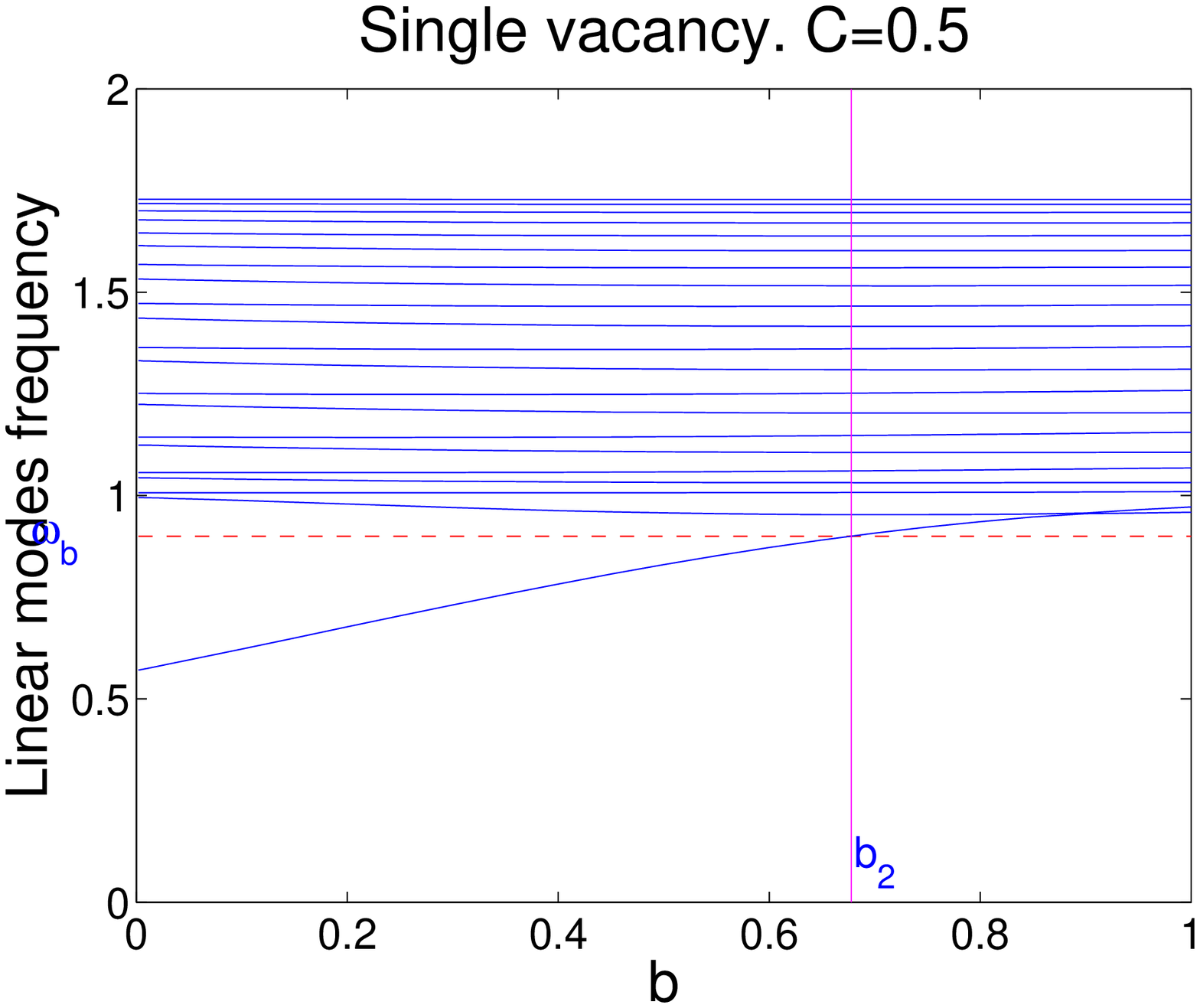} &
    \includegraphics[width=\middlefig]{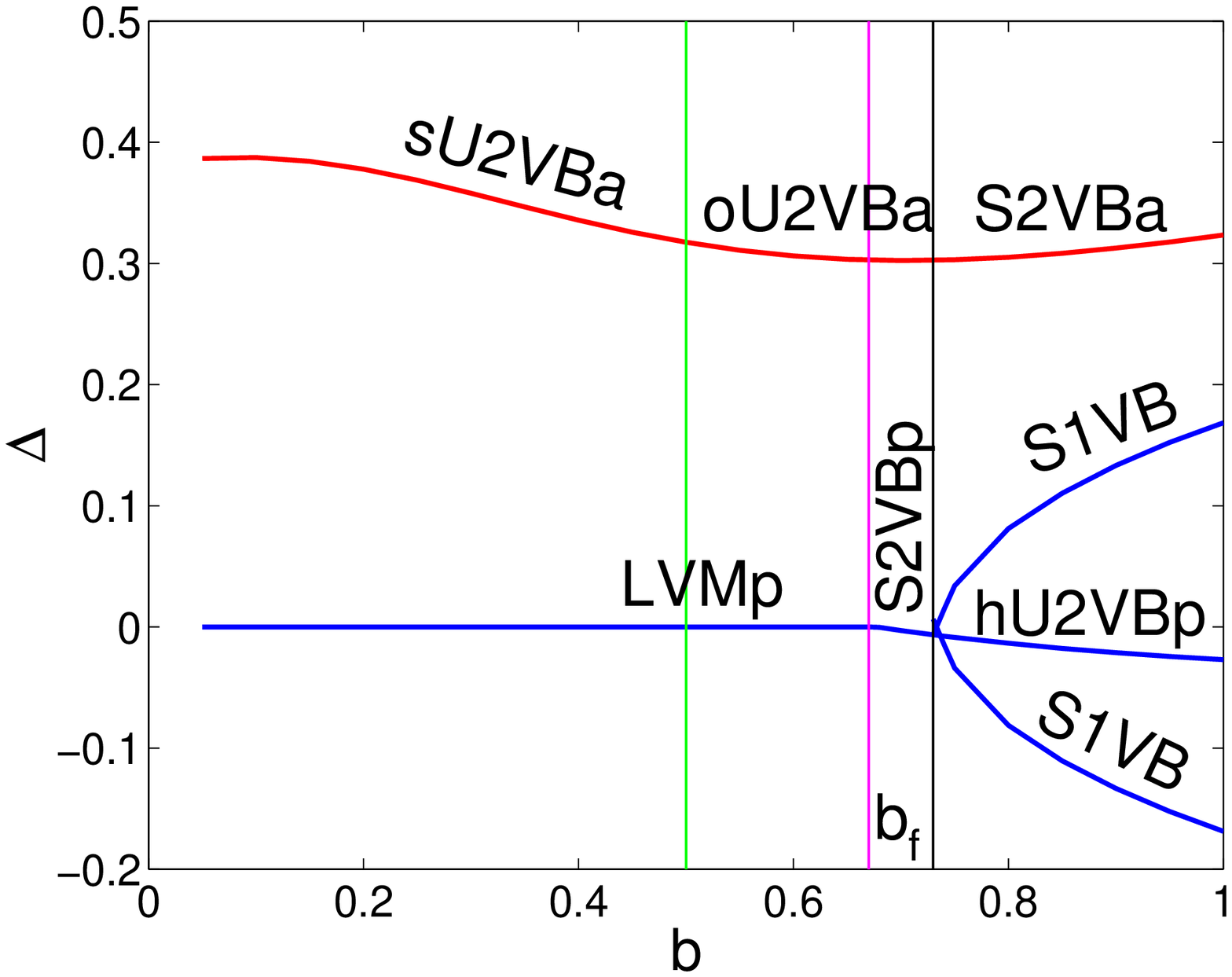}
\end{tabular}
\caption{ a) Linear spectrum. Also represented is the breather
frequency $\wb=0.9$. Note the existence of a localized linear mode.
(b) Bifurcation diagram for the nonlinear localized modes obtained
by exciting the particles adjacent to the vacancy. The bifurcation
variable $\Delta$ is the difference between the relative
displacements of the particles to the left and to the right of the
vacancy. The vertical lines indicate bifurcation points. See the
text for the breather codes.} \label{fig:tower1}
\end{center}
\end{figure}

Fig.~\ref{fig:tower1}a shows the dependence of the linear
frequencies with respect to $b$. Note that the minimum linear
frequency corresponds to a linear localized mode which we call
linear vacancy mode (LVM). On the other hand, there are three types
of VB: 1--site vacancy breathers (1VB), which consists of a single
excited particle adjacent to the vacancy; and 2--site vacancy
breathers, obtained by exciting the two particles adjacent to the
vacancy, which can vibrate in-phase (2VBp) or anti-phase (2VBa).
Fig.~\ref{fig:tower1}b shows the bifurcation pattern for these VBs.
A letter at the beginning of the abbreviation specifies  the
breather stability: {\em S} stable, {\em hU} harmonically unstable,
{\em sU} subharmonically unstable or {\em oU} oscillatory unstable.
For example sU2VBa means a subharmonically--unstable, two--site,
vacancy breather in anti--phase.

For weak interaction ($b$ large) there exist  stable one-site VBs
at both sides of the vacancy as stated by the theory~\cite{FW98}.
As $b$ decreases the S1VB disappears through an inverse pitchfork
bifurcation to two unstable two-site VB in phase (hU2VBp) which
becomes stable (S2VBp). Further decrease of $b$ brings about the
annihilation of the S2VBp when the breather frequency coincides
with the frequency of the LVM with the same profile.

The two--site VB in antiphase is stable (S2VBa) for the largest
values of $b$ represented in Fig.~\ref{fig:tower1}b. When the
interaction increases ($b$ decreases) this 2VB becomes oscillatory
unstable (oU2VBa), and finally subharmonically unstable (sU2VBa).

From this bifurcation diagram we can deduce that the stability of
the VBs is correlated to the existence of a kinetic energy threshold
to move the vacancy, in the interval $b\in (0.5,0.8)$. Comparison of
Fig. \ref{fig:Kmin} and \ref{fig:tower1}b suggests that 
 {\em the
existence of a subharmonically or harmonically unstable, 2-site,
vacancy breather is a necessary condition in order to get an optimum
vacancy mobility (no kinetic energy threshold)}.

Generally speaking, any approaching breather with any kinetic energy is able to move the
vacancy if at least one the two 2VBs exists and is either subharmonically or
harmonically unstable. We cannot expect, however, an exact agreement between the 
numerically exact bifurcation values for a given frequency with the observed changes in behavior for the simulations.
The reason for that is that a moving breather is obtained by perturbation with an asymmetric mode.
Therefore, it is no longer an exact solution of the dynamical equations, its frequency is shifted and it is
not unique. The bifurcation values for other frequencies will change and interaction of phonons emitted by
the MB are expected to have an (unknown) influence, hence the probability analysis of the simulations outcome.
However, the only important exception to the first assessment in this paragraph in the region $b\in (0.73,0.8)$ is
 noteworthy. 
  The role of the unstable vacancy breather is probably to act as an intermediate structure with
high oscillations for the particles nearest to the vacancy, which, being unstable, leads to one of them changing place. 
Clearly the stable S2VB does not play that role and its existence in the vicinity of that region seems to be the reason for
that discrepancy --proximity in $b$
for a $\wb=0.9$ also means proximity in $\wb$ for the a given value of $b$, which can be excited by the bundle of
the MB frequencies--.

Therefore our conclusion is that vacancy mobility is governed, for a
particular choice of the substrate potential, by the strength of the
interaction between the particles adjacent to the vacancy, which is
correlated to the existence and stability properties of vacancy
breathers. This correlation manifests in two different ways.
Firstly, the interaction has to be strong enough to move the vacancy
forward, or equivalently, there must exist no linear vacancy modes.
Secondly, a threshold value of the translational energy of the
incident MB in order to move the vacancy does not exist if an
 harmonically or subharmonically
  unstable VB exists.

Another choice of the interaction potential would alter the values
of $b$ which separate the different regimes. However, the
correlation between the vacancy mobility and the strength of the
potential would be similar.

\section{Double vacancy}

\begin{figure}
\begin{center}
    \includegraphics[width=\singlefig]{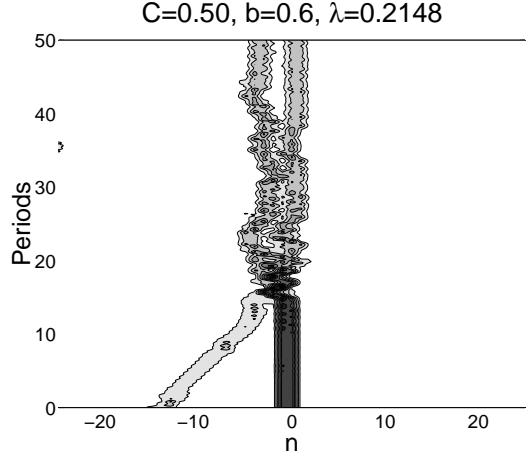} 
\caption{Energy density plot for the interaction moving
breather--double vacancy, in the vacancy splitting regime. The
particle to the right of the vacancies is located at $n=0$.}
\label{fig:split}
\end{center}
\end{figure}

\begin{figure}
\begin{center}
\begin{tabular}{cc}
    \includegraphics[width=\middlefig]{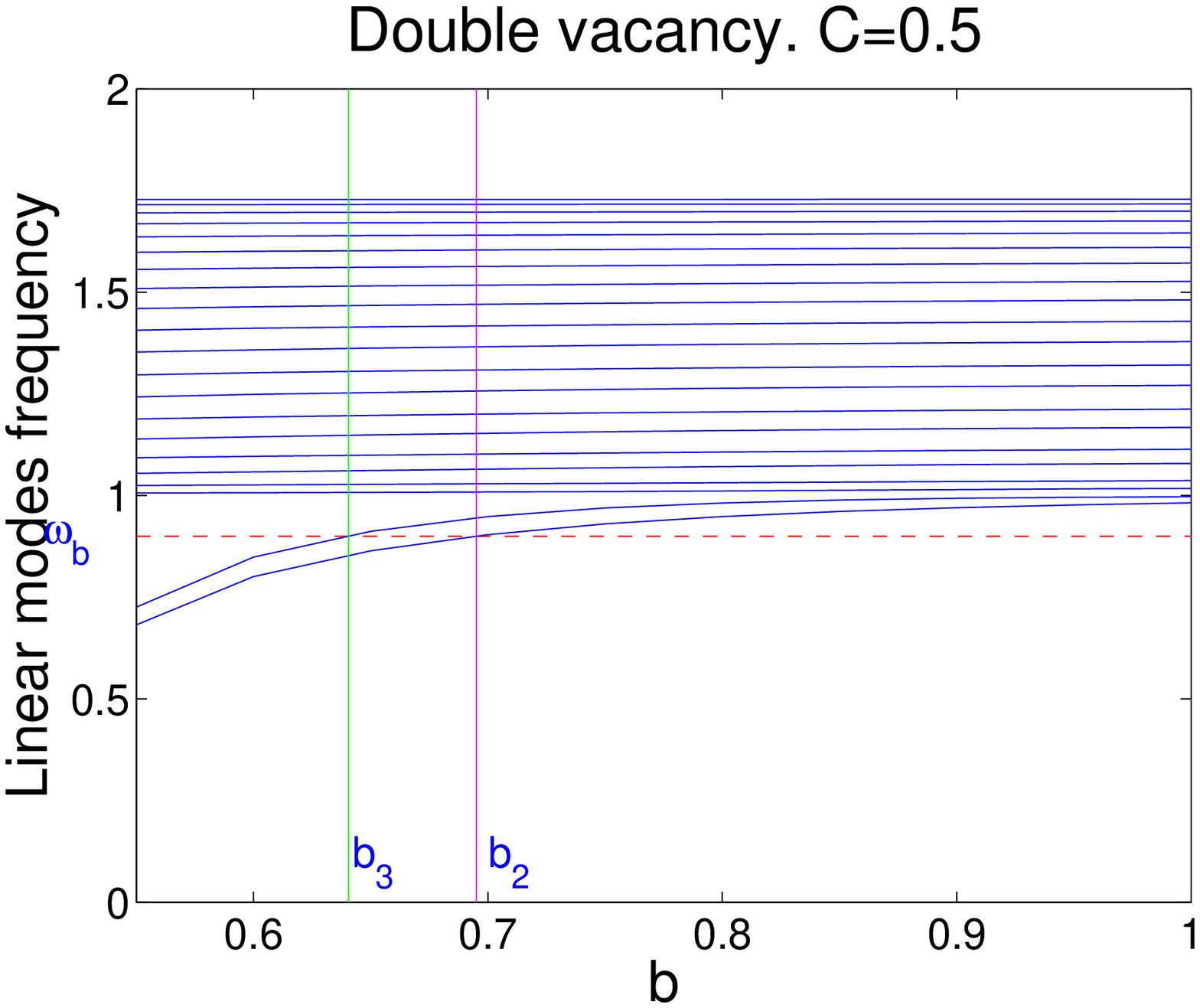} &
    \includegraphics[width=\middlefig]{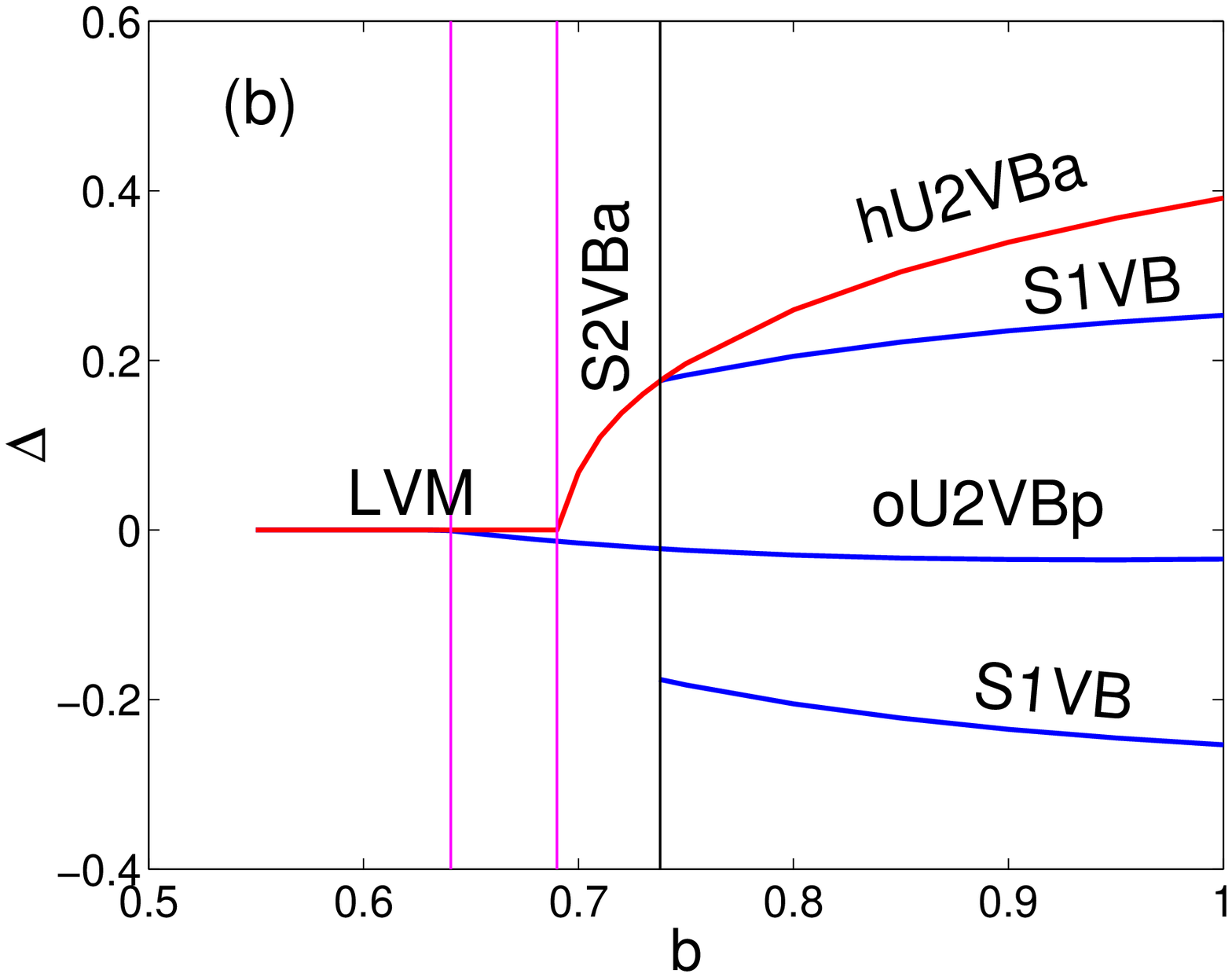}
\end{tabular}
\caption{Double vacancy case: a) Linear spectrum. The breather
frequency $\wb=0.9$ is also represented. Note the existence of two
localized linear modes. (b) Bifurcation diagram for the nonlinear
localized modes obtained exciting the particles adjacent to the
double vacancy. The bifurcation variable $\Delta$ is the difference
between the relative displacements of the particles to the left and
to the right of the double vacancy. The vertical lines indicate
bifurcation points. See text for the breather codes.}
\label{fig:tower2} \end{center}
\end{figure}

\begin{figure}
\begin{center}
    \includegraphics[width=\singlefig]{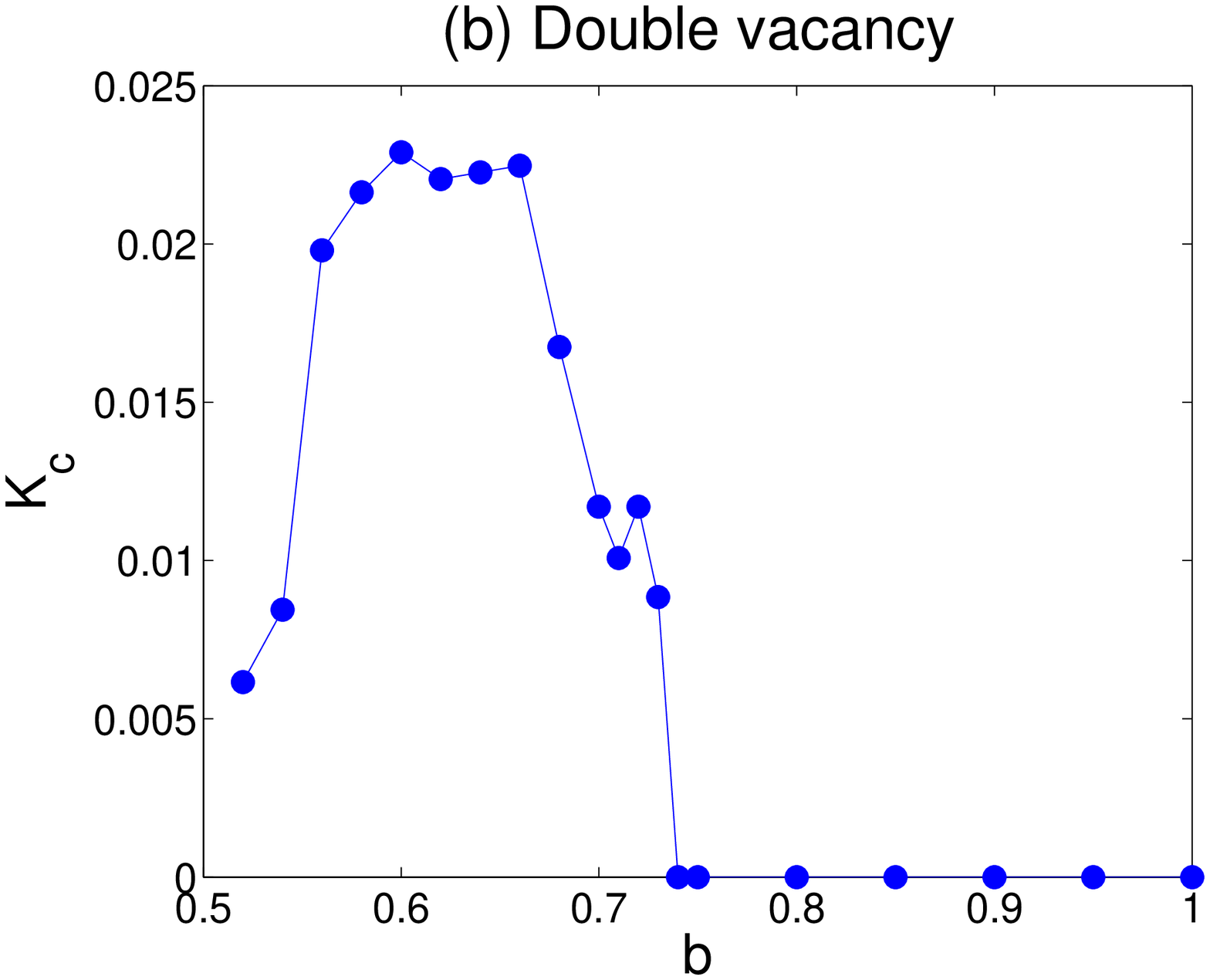} 
\caption{Minimal translational energy ($K_{min}$) to move a double
vacancy.} \label{fig:Kmin2} \end{center}
\end{figure}

We have studied the mobility of a double vacancy (two empty
neighboring wells of the substrate potential) to test our
conjecture. The outcome is similar to the single vacancy case except
for the fact that the left vacancy never moves forwards. Thus, if
the right vacancy moves forwards, the double vacancy splits in two
(see Fig. \ref{fig:split}). Fig.~\ref{fig:tower2} shows the linear
spectrum and the bifurcation diagram for the VB in this case. In
these figures the lowest value of b is approximately 0.5 since there
are no equilibrium states for $b\lesssim 0.5$. This result also
implies the non existence of double vacancies equilibrium states for
a harmonic interaction potential. As b increases, we observe
two--site VBs emerging from linear localized modes, and afterwards
an inverse pitchfork bifurcation similar to the one described for a
single vacancy. There is not any harmonically or subharmonically
unstable VB for $b<0.75$. Therefore, we should expect that there is
a minimum value for the kinetic energy of the incident MBs able to
move the double vacancy. This is confirmed by numerical computations
as Fig.\ref{fig:Kmin2} shows.

\section{Summary}

We have observed numerically the interaction of moving breathers with vacancies
in the simplest, physically consistent model. We have found that the breathers
can be reflected, transmitted and trapped, both by the single and the double
vacancy. The single vacancy can move forward, backwards and remain and rest,
but the double vacancy cannot move forward, instead it can split into two
single vacancies, one at rest and the other moving forwards. This phenomenology
is very different from the continuous analogue, where the breather is always
transmitted and the vacancy (anti-kink) always moves backwards.

The vacancy migration is strongly correlated with the existence and
stability of vacancy breathers, which we have studied in detail. In
order to move the vacancy the interaction has to be strong enough
and the energy of the incident MB has to be above a threshold. This
threshold disappears approximately around bifurcation points where
vacancy breathers become unstable.

We think that the study of the properties of breathers next to a defect may
help to understand the mobility of other class of point defects such as
interstitials, which might be related with recent defect migration observed
experimentally in ion-irradiated silicon.

\section{Acknowledgments}

The authors acknowledges sponsorship by the Ministerio de Educaci\'{o}n
y Ciencia, Spain, project FIS2004-01183. They also acknowledge JC
Eilbeck and F Palmero for useful discussions.

\newcommand{\noopsort}[1]{} \newcommand{\printfirst}[2]{#1}
  \newcommand{\singleletter}[1]{#1} \newcommand{\switchargs}[2]{#2#1}

\end{document}